\documentclass[fleqn,12pt]{wlscirep}
\usepackage[utf8]{inputenc}
\usepackage[T1]{fontenc}
\title{Forman--Ricci Curvature for Irregular Convex Mosaics}

\author[1]{Abhyudaya Gupta}
\author[2]{Sayak Mukherjee}
\author[3]{Kuldeep Saha}
\affil[1]{Pangea Society, India (abhyudaya.gupta.ag@gmail.com)}
\affil[2]{Kolkata, India (sayakm.2311@gmail.com)}

\affil[3]{IAI TCG CREST Kolkata, India 
(kuldeep.saha@tcgcrest.org)}

\begin{abstract}
Forman has defined a discrete version of the Ricci curvature on Riemannian manifolds, known as the Forman--Ricci curvature. The Forman--Ricci curvature has found significant applications in several pattern-recognition problems occurring in natural sciences. Domokos and Langi, on the other hand, have defined a notion of irregularity for convex mosaics, which has also found remarkable applications to the geological problem of fractures in rocks. We define a modification of the classical Forman--Ricci curvature for irregular convex mosaics and demonstrate how they can be used to distinguish between various fractures or cracking patterns appearing in nature. 
\end{abstract}
\begin{document}

\flushbottom
\maketitle
%
%
\thispagestyle{empty}

\section*{Introduction}

\emph{Mosaic} (or \emph{tiling}) is an ancient field of study in geometry and architecture. It is a natural way to fill surfaces of $3$-dimensional objects by smaller blocks of prescribed polygons. The theoretical study of mosaic patterns has been an intriguing field of mathematics. Moreover, it appears in the context of many other natural sciences, as it is one of the most abundant types of data available around us. From cracked surfaces on a rock to tectonic fault lines on planetary landscapes, mosaics are everywhere in nature. In a recent article, Domokos and Langi \cite{dom2} introduced the notion of a \emph{symbolic plane} to compare various convex irregular mosaics. Here, \emph{convex} means that each building block of the mosaic is a convex polygon, and \emph{irregular} means that there might be edges in the mosaic that are not part of a straight line. For example, if the local picture near a node looks like \textbf{X}, then it is a \emph{regular node}. A node with a local picture like \textbf{Y}, will be an \emph{irregular node}. Using the notion of symbolic plane, Domokos etal. \cite{dom1} differentiated various types of 2D and 3D mosaics that occur in geophysical studies of rocks and earth surfaces. They demonstrate a remarkable fact that on an overage, the most probable shape for a building block of a generic 2D convex mosaic is a $4$-gon (a hexahedron in case of a 3D mosaic). On the other hand, an important parameter for a network (i.e., a simple graph) is the \emph{Forman--Ricci curvature}\cite{For}. It is a discrete version of the classical Ricci curvature on a Riemannian manifold. The Forman--Ricci curvature has found many applications in the study of complex network systems arising from neuroscience, social networking, and molecular biology. For example, it is able to distinguish an autistic human brain from a normal one by comparing the curvature distributions of the networks derived from fMRI images \cite{jost1}. Further applications can be found in Jost et al. \cite{jost2}. In the present article, we introduce an \emph{irregular} version of the Forman--Ricci curvature of a network whenever one can make sense of a regular/irregular node in a network consisting of a weighted simple graph.   

\begin{figure}[ht]
\centering
\includegraphics[width=14cm]{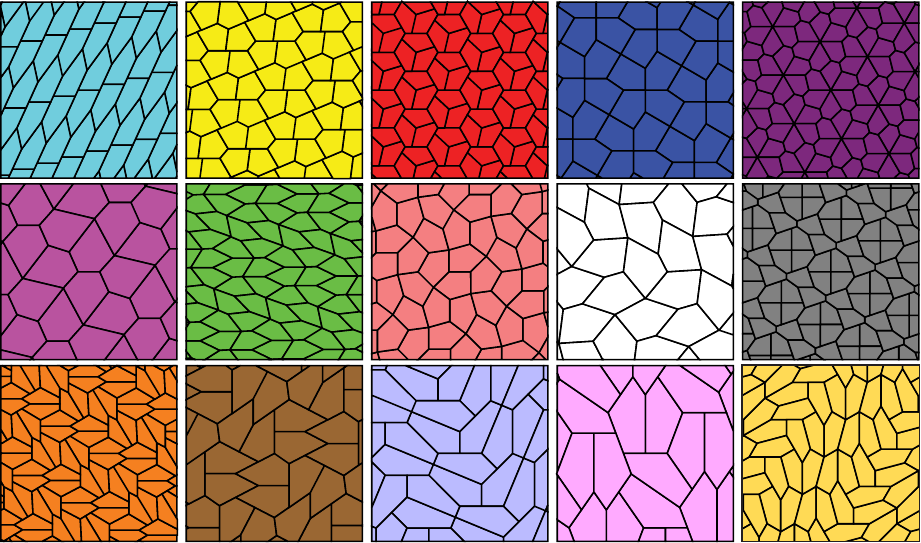}
\caption{A family of irregular convex pentagonal mosaics which can not always be distinguished by the usual Ricci-Forman curvature, but often distinguished by the regular/irregular version of it. Image taken from https://blogs.ams.org/blogonmathblogs/2015/09/07/theres-something-about-pentagons/}
\label{fig0}
\end{figure}

\section*{2-dimensional mosaics and their discrete curvature}

In this section, we recall some preliminary mathematical concepts and define the notion of an \emph{irregular} version of the Forman-Ricci curvature in the context of convex, irregular mosaics on a two-dimensional Euclidean plane. Much of these notions can be extended to a general orientable surface.

\subsection*{Convex irregular mosaics on two dimensional plane} As described in the works of Domokos et al.\cite{dom1}, fragments and cracks on a planar surface can be represented by the \emph{cells} of a convex mosaic. In a convex mosaic, each enclosed region, a $2$-dimensional cell or \emph{face}, is bounded by a convex polygon. A \emph{node} or \emph{vertex} is a zero-dimensional cell and an \emph{edge} is a $1$-dimensional cell. Let $\bar{v}$ denote the average number of vertices of the polygons, and let $\bar{n}$ denote the average number of vertices that meet at a node (i.e. average \emph{degree}of a node). Domokos et al used these two parameters for distinguishing various convex mosaics coming from geological sources. the $2$-plane parametrized by $\bar{n}$ and $\bar{v}$ is called the \emph{the symbolic plane}. The work of Domokos et al PNAS, also introduces the notion of \emph{irregularity } in a convex mosaic. Each node in a convex mosaic has only two types of edges attached to it. 

\begin{enumerate}

    \item[(a)] Edges coming from intersecting straight lines, called \emph{straight edges}, and
    \item[(b)] Edges not coming from intersecting straight lines, called \emph{skewed edges}.     
    
\end{enumerate}

\noindent A node is called \emph{regular}, if it is not contained in a skewed edge. A node containing skewed edges will be called \emph{irregular}.

\begin{figure}[ht]
\centering
\includegraphics[width=14cm]{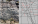}
\caption{This picture, taken from Domokos etal. \cite{dom1}, gives an example of irregular mosaic pattern on a cracked rock surface. The red edges on the right hand diagram are the irregular ones. Any node belonging to these irregular edges is an irregular node.}
\label{fig1}
\end{figure}

\noindent Let $N_R$ and $N_I$ denote the number of regular and irregular nodes, respectively, in a mosaic. Domokos et al defined $p = \frac{N_R}{N_R + N_I}$ as the \emph{regularity factor} of the mosaic. A regular (irregular) mosaic will have $p = 1$ ($p = 0$). The following relation between $\bar{n}$, $\bar{v}$ and $p$ was proved by Domokos-Langi.

$$ \bar{v} = \frac{2 \bar{n}}{\bar{n} - p - 1}$$. 

\noindent The above relation helps compare irregularity factors between various convex mosaics. We note that for a general irregular convex mosaic, $\bar{n}$ is defined as the \emph{average regular degree} of a node in the mosaic. Here, \emph{regular degree} $d_R(u)$ of a node $u$ is the number of straight edges attached to it. Similarly, the \emph{irregular degree} $d_I(u)$ is the number of skewed edges attached to $u$. 

\begin{figure}[ht]
\centering
\includegraphics[width=12cm]{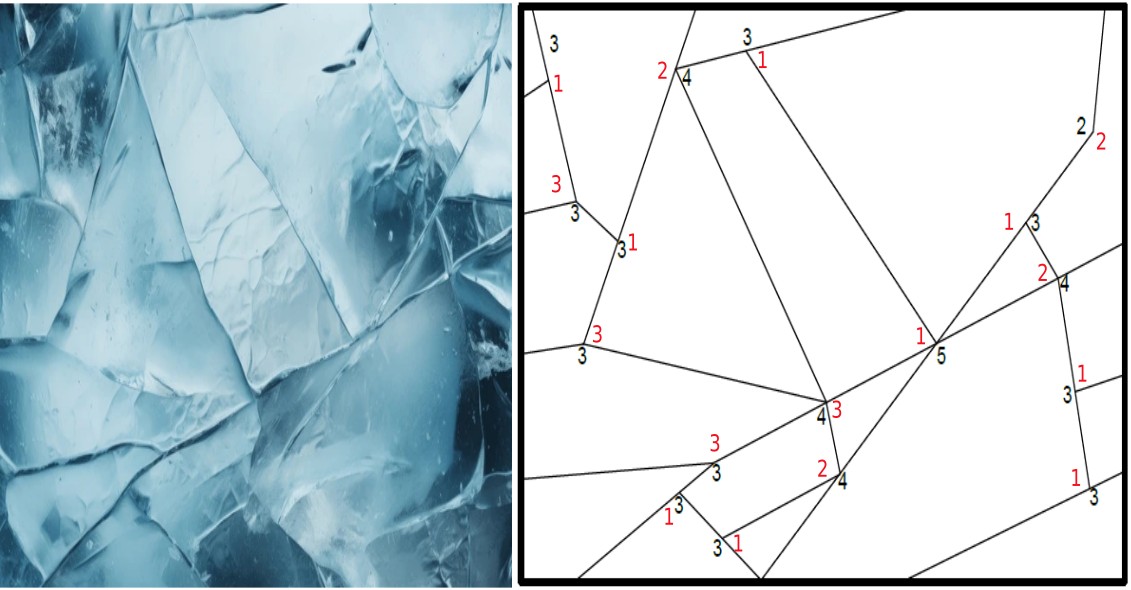}
\caption{Plotting classical nodal degree (numbers in black) and irregular nodal degree (numbers in red) for a mosaic on a cracked ice surface.}
\label{fig2}
\end{figure}

\subsection*{Edge-based Forman--Ricci curvature} Forman [] introduced a discrete version of the Ricci curvature in Riemannian geometry, called the \emph{Forman--Ricci curvature}. For a detailed description of the background and applications of the Forman--Ricci curvature, we refer to Jost []. We will only consider the Forman--Ricci curvature for an edge. Let us label the nodes in a finite network from $1$ to $k$. For $i,j \in \{1,2,\dots,k\}$, let $[i,j]$ denote an edge joining the vertices $i$ and $j$. The Forman--Ricci curvature $c_{RF}$ of $[i,j$ id defined as follows.

$$ c_{RF}([i,j]) = deg(i) + deg(j) - 4$$

\noindent Here, $\deg(u)$ denotes the usual degree of a node $u$.

\subsection*{Modified Forman--Ricci curvature for a convex mosaic} \label{meth1}

For an irregular convex mosaic $\mathcal{M}$ on a $2$d plane, let $G(\mathcal{M})$ denote the corresponding simple, planar graph. As before, we label the nodes from $1$ to $k$. Recall that $d_R(u)$ denotes the regular degree of a node $u$ in $\mathcal{M}$. We define two modifications of the Forman--Ricci curvature for an edge $[i,j]$ : the \emph{regular curvature} $c^{reg}_{RF}$ and the \emph{irregular curvature}
$c^{irreg}_{RF}$.
$$c^{reg}_{RF} = d_R(i) + d_R(j) - 4$$
$$c^{irreg}_{RF} = d_I(i) + d_I(j) - 4$$

\subsection*{Modification for node based Forman--Ricci curvature}\label{meth2} The Forman--Ricci curvature of a node is the sum of the Forman--Ricci curvatures of the incident edges. In particular, for a node $u$ in a network we have $$c_{RF}(u) = \frac{1}{deg(u)}\sum_{w \sim u} c_{RF}([u,w]) = deg(u) + \sum_{w \sim u} \frac{deg(w)}{deg(u)} - 4$$. 

\noindent The modified node-based regular curvature for a generic convex mosaic is then defined by

$$c^{reg}_{RF}(u) = \frac{1}{deg(u)}\sum_{w \sim u} c^{reg}_{RF}([u,w]) = d_R(u) + \sum_{w \sim u} \frac{d_R(w)}{deg(u)} - 4$$.

\noindent The average or mean node-based regular curvature is defined as $\bar{c}_R = \frac{\sum_{u} c^{reg}_{RF}(u)}{N}$, where $N$ is the total number of nodes. Similarly, one can define the average node-based irregular curvature as $\bar{c}_I = \frac{\sum_{u} c^{irreg}_{RF}(u)}{N}$.

\subsection*{Pair of distributions for node based $c^{reg}_{RF}$ and edge based $c_{RF}$} In Domokos et al, the key idea to compare irregular convex mosaics was to plot their corresponding $(\bar{n},\bar{v})$ coordinate on the symbolic plane. This approach measured the effect of irregular nodes by comparing the average number of nodes in a face ($\bar{v}$) and the average regular degree of a node ($\bar{n}$). We propose an analogous approach to measure irregularity by comparing pairs of curvature distributions : one for node-based regular curvature $c^{reg}_{RF}$, and the other for edge-based Forman--Ricci curvature $c_{RF}$. Moreover, the corresponding mean curvatures $\bar{c}_R$ and $\bar{c}_I$ can be plotted similar to the symbolic plane.

\subsection*{Relation with the weighted version of Forman--Ricci curvature} Consider an weighted network with weights associated to both edges and vertices. For an edge $e$ in the network, joining vertices $v_1$ and $v_2$, the weighted version of the Forman--Ricci curvature is given by the following formula \cite{jost2}.

$$c^w_{RF} = - w_e\left(\frac{2w_{v_1}}{w_e} + \frac{2w_{v_2}}{w_e} - \sum_{e_{v_1} \sim e, e_{v_2} \sim e}\left[\frac{w_{v_1}}{\sqrt{w_e w_{e_{v_1}}}} + \frac{w_{v_2}}{\sqrt{w_e w_{e_{v_2}}}}\right]\right)$$.

\noindent Here, $w_e$ denotes the weight of the edge $e$, $w_{v_1}$ and $w_{v_2}$ denote the weights of the nodes $v_1$ and $v_2$, respectively. $e_{v_1} \sim e$ and $e_{v_2} \sim e$ denote the set of edges incident on nodes $v_1$ and $v_2$, respectively, excluding the edge $e$.

\noindent The definition of regular curvature for the network associated with irregular convex mosaics can be interpreted in terms of the weighted Forman--Ricci curvature once we make the following modification on the underlying network. Define all weights on the edges and vertices to be $1$ and the summation in the formula is taken over $e_{v_1}$ and $e_{v_2}$ such that both $e_{v_1}$ and $e_{v_2}$ are straight edges. Thus, the regular version of Forman--Ricci curvature also makes sense for an weighted irregular convex mosaic.

\section*{Computations on some surface fractures in nature} 
\begin{figure}[ht]
\centering
\includegraphics[width=14cm]{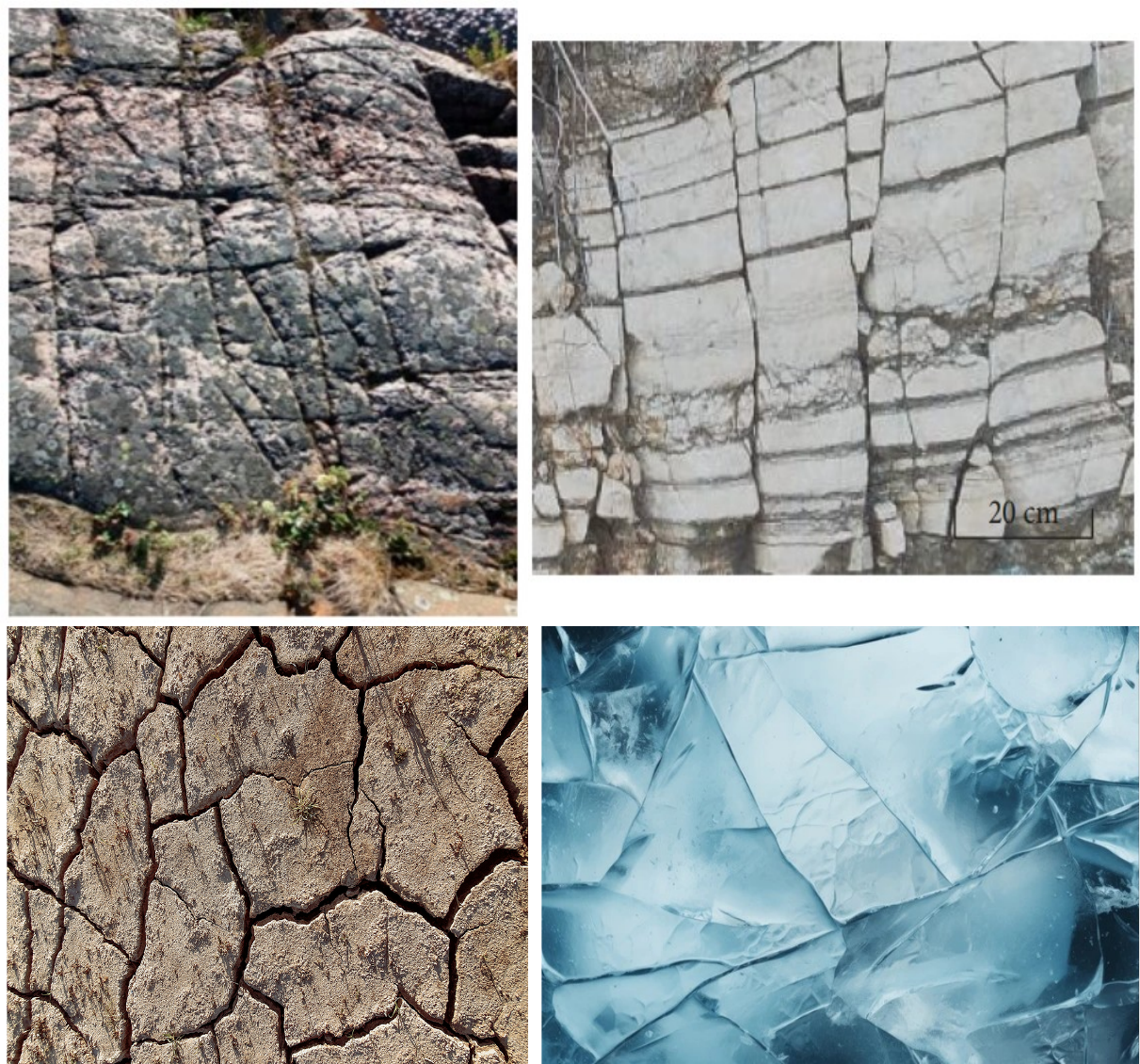}
\caption{The north-west picture is taken from Domokos et al.\cite{dom1} and it shows mosaics on a granite surface. The north-east picture, taken from Zhang et al.\cite{zll}, shows natural fracture network of outcrops. The south-west picture shows desiccation cracks due to drought which enhances the release of greenhouse gases \cite{cutts} (photo credit: José Ignacio Pompé/Unsplash). The south-east picture shows a local crack pattern in a glacier. All of these patterns can be approximated by irregular mosaics.}
\label{hfigc}
\end{figure}

We now demonstrate how the modified Forman--Ricci curvature can help distinguish between various mosaic patterns appearing in the real world. We focus on four such examples (See Figure \ref{hfigc}). For simplicity, we restrict ourselves to the standard unweighted case. Even for these unweighted cases, we see significant variations in the modified curvature distributions.

\begin{figure}[htbp]
\centering
\includegraphics[width=14cm]{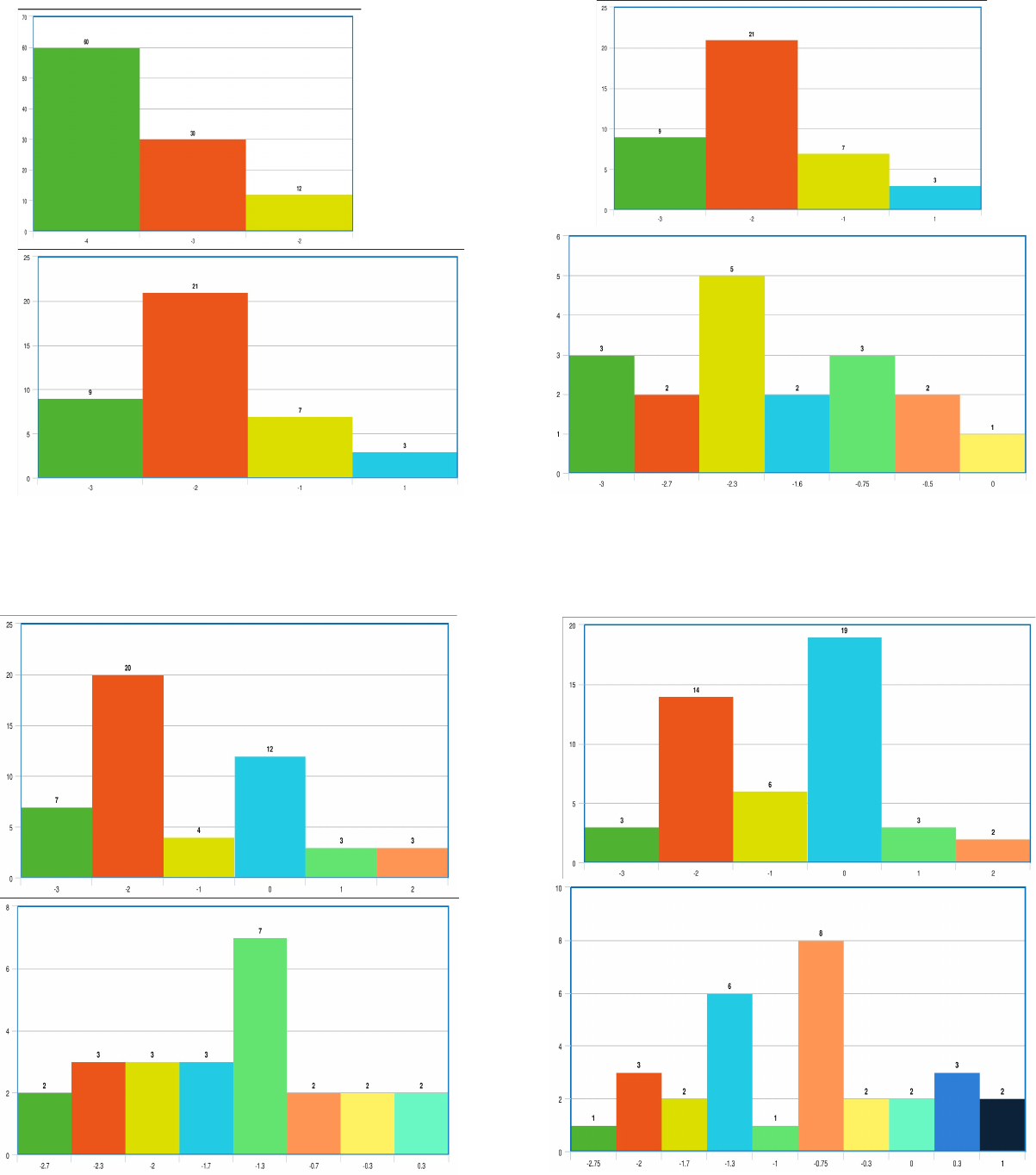}
\caption{The distributions of \emph{edge based irregular} Forman--Ricci curvature and \emph{node based irregular} Forman--Ricci curvature corresponding to the real-world mosaics shown in Figure \ref{hfigc}, in the same clockwise order. In each of the four blocks (north-west, north-east, south-west and south-east), the upper histogram shows distribution of \emph{edge based irregular} Forman--Ricci curvature, and the lower histogram shows distribution of \emph{node based irregular} Forman--Ricci curvature. The X-axis plots the curvature values (increasing order) and the Y-axis plots their frequencies.}
\label{hfigc2}
\end{figure}

Figure \ref{hfigc2} demonstrates the distributions of \emph{edge based irregular} Forman--Ricci curvature and \emph{node based irregular} Forman--Ricci curvature corresponding to the mosaic patterns shown in Figure \ref{hfigc}, in the same clockwise order. The distributions clearly distinguishes the four mosaic patterns. For large-scale samples, such distributions can be approximated by various functions (eg. via regression methods) and the notion of their "differences" can be made mathematically precise via various \emph{metrics} on the space of functions (eg. the \emph{Wasserstein metric}). 

\vspace{0.25cm}

\noindent We explain the methodology in two major steps. First, we describe our method to process the raw image of natural fractures and convert them into digital irregular mosaics or networks. Then, we show how to compute the Ricci--Forman curvature for these networks.

\subsection*{From raw image to irregular network} We convert a normal/natural image into a binary mask where white represents the cracks and the background / noise is black. We use the following libraries :

\begin{enumerate}
    \item OpenCV (cv2): image processing (read, grayscale, contrast, edges, morphology, line detection)
    \item NumPy (numpy): kernels and array utilities
    \item Matplotlib (matplotlib.pyplot): optional visualization only
\end{enumerate}

\noindent The following gives a rough description of the conversion process used by the authors.

\begin{enumerate}
    \item[Step 1:] The image is first loaded using OpenCV (cv2.imread), which stores it as a 3-channel RGB matrix where each pixel has intensity values from (0) to (255); this is the raw input signal for the entire pipeline.
    \item[Step 2:] The RGB image is converted to grayscale using cv2.cvtColor, cv2.COLOR\textunderscore BGR2GRAY. This reduces complexity and keeps the structural information needed for cracks. 
    \item[Step 3:] Local contrast is enhanced using CLAHE (cv2.createCLAHE(...).apply(gray)), which performs adaptive histogram equalization on small tiles and clips histogram peaks to avoid amplifying noise too much.
    \item[Step 4:] The enhanced grayscale image is denoised with Gaussian blur (cv2.GaussianBlur) so random texture and sensor noise do not create false edges.
    \item[Step 5:] Edges are extracted using Canny (cv2.Canny) which uses Canny edge detection algorithm. It first computes the gradient along every x and y coordinate and then based on where the brightness changes are the sharpest it identifies edges and creates a binary edge map.
    \item[Step 6:] The edge map is cleaned with morphology (cv2.morphologyEx) by first filling small gaps so broken lines connect (closing), and then removing tiny random noise dots (opening).
    \item[Step 7:] Crack-like line segments are detected using the probabilistic Hough transform (cv2.HoughLinesP), which identifies straight lines by mapping edge points into a parameter space defined by $\rho = xcos\theta+ysin\theta$. Only line segments that meet specified thresholds such as minimum length are kept, and then it is drawn onto a blank image.
    \item[Step 8:] The drawn crack-segment image is converted to a final binary mask with thresholding (cv2.threshold) which maps the spectrum of colors to either completely black or completely white. This essentially converts the grayscale image into a true binary mask where crack pixels are in white and everything else in black.
    \item[Step 9:] The final crack-only mask is saved to the computer with cv2.imwrite. If the image needs to be visualized with Matplotlib (matplotlib.pyplot), it creates a popup and shows the result image along with the original image.     
\end{enumerate}

\noindent The details of our experimental code is available here : \emph{https://github.com/AbhyudayaGup/Crack-Identification}. For large-scale data, a nice and efficient approach is given in \cite{yia} for image processing. A similar approach could be used for large-scale creation of digitized mosaic networks.

\subsection*{Computing modified Ricci-Forman curvature} Starting with real-world mosaic pictures, we can create a network of edges and vertices (as shown in Figure \ref{fig1} and in Figure \ref{fig2}). The next step describes how to calculate the regular/irregular nodal degrees and Forman--Ricci curvatures.

\begin{enumerate}
    \item[Step 1:] In a digitized mosaic or network, the vertices are given as 2D vectors $v_i = (x_i,y_i)$, and edges are given as pairs $[v_i,v_j] = e_{ij}$.
    
    \item[Step 2:] Given a node $v_i$, the angle between edges $e_{ij}$ and $e_{ik}$ is given by $\theta^i_{j,k} = \tan^{-1}[\frac{m_{ij} - m_{ik}}{1 + m_{ij}m_{ik}}]$. Here, $m_{lk} = \frac{y_l - y_k}{x_l - x_k}$.

    \item[Step 3:] Let For a given $v_i$, compute all $\theta^i_{j,k}$ for pairs of neighbouring vertices $v_j,v_k$. There will be $\frac{deg(v_i) (deg(v_i)-1)}{2}$ many such pairs. If $\theta^i_{j,k} \in [165^\circ, 180^\circ]$ for all $k$, then $e_{ij}$ is a regular edge. The number of such $e_{ij}$s gives the irregular degree of $v_i$ : $d_I(v_i) = deg(v_i) - d_R(v_i)$. 
    
    \item[Step 4:] Now compute the edge based and node based regular (and irregular) versions of Forman--Ricci curvatures using the formula

    $$c^{reg}_{RF}(u) = \frac{1}{deg(u)}\sum_{w \sim u} c^{reg}_{RF}([u,w]) = d_R(u) + \sum_{w \sim u} \frac{d_R(w)}{deg(u)} - 4$$

    \item[Step 5:] Plot the distributions of the corresponding Forman--Ricci curvatures.  
\end{enumerate}

\noindent Our sample code for computing the Forman--Ricci curvature can be found here : \emph{https://shorturl.at/tTE8l}. For the fracture patterns in Figure \ref{hfigc}, the computed values of the parameters are given below.

\begin{itemize}
    \item[North-West] Edge-based $c^{irreg}_{RF}$, values(frequency): $-4$ ($60$), $-2$ ($12$), $-3$ ($30$).
    
      Node-based $c^{irreg}_{RF}$, values(frequency): $-4$ ($30$), $-3.75$ (20), $-3.5$ ($8$), $-3.25$ ($2$), $-2$ ($15$), $-1.5$ ($2$).

    \item[North-East] Edge-based $c^{irreg}_{RF}$, values(frequency): $-2$ ($21$),$-3$ ($9$),$-1$ ($7$),$1$ ($3$).

    Node-based $c^{irreg}_{RF}$, values(frequency): $-2$ ($12$),$-2.3$ ($5$),$-1.6$ ($2$),$-3$ ($3$),$-2.7$ ($2$),$-0.5$ ($2$),$-0.75$ ($3$),$0$ ($1$). 

    \item[South-West] Edge-based $c^{irreg}_{RF}$, values(frequency): $-2$ ($20$),$0$ ($12$),$-1$ ($4$),$1$ ($3$),$-3$($7$), $2$ ($3$).

    Node-based $c^{irreg}_{RF}$, values(frequency): $-2$ ($3$),$-1.3$ ($7$),$-1.7$ ($3$),$-0.7$ ($2$),$0.3$ ($2$),$-0.3$ ($2$),$-2.7$ ($2$),$-2.3$ ($3$). 

    \item[South-East] Edge-based $c^{irreg}_{RF}$, values(frequency): $0$ ($19$),$-2$ ($14$),$-1$ ($6$),$1$ ($3$),$-3$($3$), $2$ ($2$).

    Node-based $c^{irreg}_{RF}$, values(frequency): $0$ ($2$),$-1.3$ ($6$),$-1.7$ ($1$),$-0.75$ ($8$),$0.3$ ($3$),$-0.3$ ($2$),$-2.75$ ($1$),$-2$ ($3$), $1$ ($2$), $-1$ ($1$). 
    
\end{itemize}

\noindent The corresponding curvature distributions are plotted in Figure \ref{hfigc2}.

\section*{Extension to higher dimensional mosaics/cracks} The definition of irregular/regular curvature can be extended to higher dimensions by generalising the notion of \emph{irregular edges} to \emph{irregular $p$-cells} ($p \geq2$). For simplicity we describe here the generalization to $3$-dimensional mosaics or fractures. The extension to an arbitrary CW complex of higher dimension essentially follows the same idea, but needs more technical definitions from combinatorial/discrete topology. For details of Forman--Ricci curvature of a general $p$-cell we refer to the original paper of Forman \cite{For}.

\noindent A $3$-dimensional mosaic or fracture on a solid object divides it into finitely many $3$-dimensional polyhedrons separated by $2$-dimensional planes. If one considers the union of all separating planes of the mosaic, one obtains a \emph{$2$-complex} that encodes all the data needed to describe the fracture. This is similar to the $2$-dimensional case, where the underlying graph/ network described the mosaic.

A $p$-simplex is defined as the convex hull of $p+1$ independent points $\{v_1, v_2, \dots, v_{p+1}\}$ in $\mathbb{R}^{p+1}$. The boundary of a $p$-simplex is the union of $p$ many $()p-1)$-simplices, where the $i$th $(p-1)$-simplex is the convex hull of $\{v_1, v_2, \dots, v_{i-1}, v_{i+1},\dots, v_{p+1}\}$, for$1 \leq i \leq p+1$). If a $(p-1)$-cell $\sigma^{p-1}$ is a boundary face of a $p$-cell $\alpha^p$, then we denote it by $\alpha > \sigma$, or $\sigma < \alpha$.

\noindent Two $p$-simplices $\alpha^p$ and $\tilde{\alpha}^p$ are called \emph{parallel} (denoted by $\alpha || \tilde{\alpha}$) if exactly one of the following two holds.

\begin{enumerate}
	\item There is a $(p+1)$-simplex $\beta^{p+1}$ such that $\alpha < \beta$ and $\tilde{\alpha} < \beta$.
	\item  There is a $(p-1)$-simplex $\gamma^{p-1}$ such that $\gamma < \alpha$ and $\gamma < \tilde{\alpha}$.
\end{enumerate}

\noindent Let $X$ be a simplicial complex. The $p$th Forman--Ricci curvature of a $p$-simplex $\alpha^p \subset X$ is defined as follows. 

\vspace{0.25cm}

$\mathcal{F}_p(\alpha) = \# \{(p+1)-\textrm{simplex} \ \ \beta, \beta > \alpha \} + \# \{(p-1)-\textrm{simplex}  \ \ \gamma, \gamma < \alpha\} - \# \{p-\textrm{simplex} \ \ \tilde{\alpha}, \alpha || \tilde{\alpha}\}$. 

\vspace{0.25cm}

\noindent For an weighted version of the above definition see \cite{For}.

\

Thus, for a $3$-dimensional mosaic, the $2$-dimensional curvature of a $2$-dimensional face/domain $\alpha^2$ as $\mathcal{F}_2(\alpha)= \# \{3-\textrm{dimensional polyhedron}  \ \ \beta, \beta > \alpha \} + \# \{\textrm{edge} e, e < \alpha\} - \# \{2-\textrm{faces} \ \ \tilde{\alpha}, \alpha || \tilde{\alpha}\}$. The corresponding $1$-dimensional curvature of an edge $e$ would be $\mathcal{F}_1(e) = \# \{2-\textrm{faces}  \ \ \beta, e<\beta\} + 2 - \# \{\tilde{e}, e|| \tilde{e}\}$.

\begin{figure}[ht]
	\centering
	\includegraphics[width=7cm]{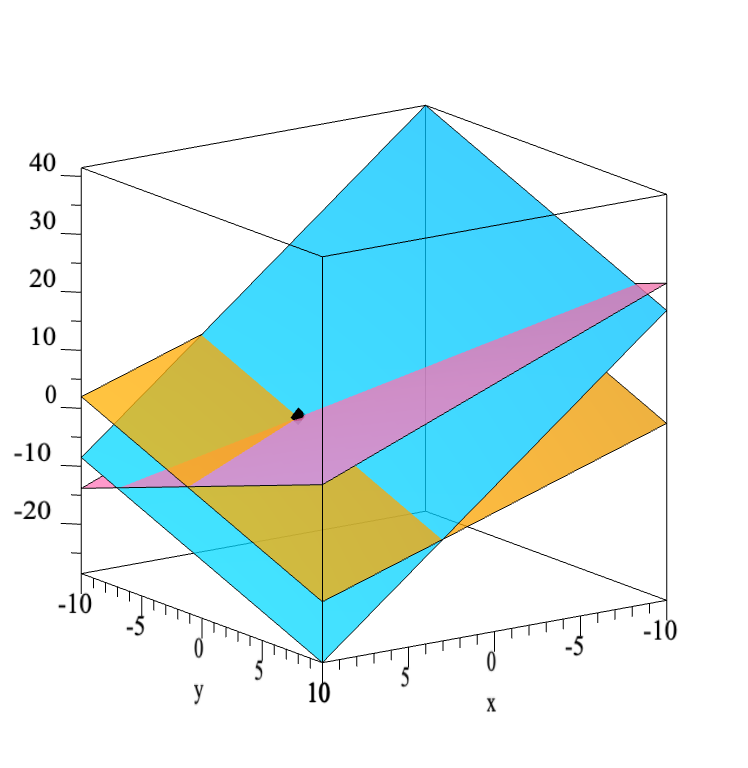}
	\caption{A generic picture of planes intersecting in $\mathbb{R}^3$.}
	\label{fig6}
\end{figure}

\noindent In a convex $3$-dimensional mosaic, the notion of a regular edge remains the same as in the $2$-dimensional case. However, unlike in dimension $2$, now $e$ can be in the boundary of more than $2$-faces. We call a $2$-face $F$, \emph{regular} if it satisfies the following conditions.

\begin{enumerate}
	
	\item If $e < F$, then $e$ is a regular edge.
	
	\item For each edge $e$ in the boundary of $F$, there is a $2$-face $\tilde{F}$ ($\neq F$) such that $e = F \cap \tilde{F}$ and the angle between $F$ and $\tilde{F}$ is $180^\circ$. One could visualise this similar to the $2$-dimensional case. In the $2$-dimensional case, the feature of regularity was based on the notion of a skewed edge and a straight edge. For example, if a neighborhood of an irregular node looks like \emph{Y} and a neighborhood of a regular node looks like \emph{X}, then $Y \times [0,1]$ and $X \times [0,1]$, as sets in $\mathbb{R}^3$, will represent neighbourhoods of an irregular edge and a regular edge, respectively.  
\end{enumerate}

\vspace{0.15cm}

\noindent Thus, in a $3$D mosaic the formula for regular Forman curvatures could be defined in the following way.

\begin{itemize}
	\item $\mathcal{F}^{reg}_1(e) = \# \{\textrm{regular 2-faces} \ \ F,  F > e \} - \# \{\textrm{regular edge \ \ }\hat{e}, \hat{e} || e \}+ 2$.  
	\item $\mathcal{F}^{reg}_2(F) = 2 + \# \{\textrm{regular edge} \ \ e,  e< F \} - \# \{\textrm{regular 2-face} \ \ \hat{F}, \hat{F} || F \}$.
\end{itemize}

\begin{figure}[ht]
	\centering
	\includegraphics[width=12cm]{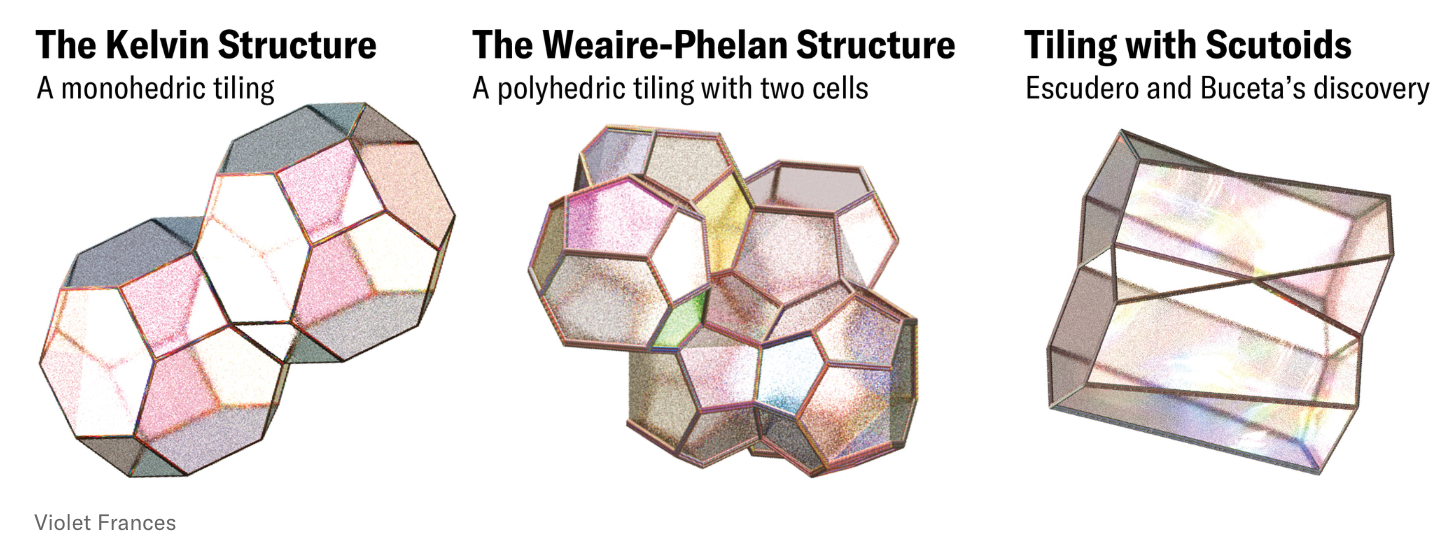}
	\caption{Examples of irregular 3D mosaics. Image source : https://sl1nk.com/bekdi7c}
	\label{fig7}
\end{figure}

\noindent Similarly, one can define the notion of \emph{irregular} curvature in this context by counting only edges and faces which are not regular. 

\section{From curvature to homology} 

One can find $3$-dimensional solid objects in nature representing various types of topology. The most trivial among them is the $3$-ball. There are three other major types.

\begin{itemize}
	\item Doughnut like object (eg. coffe mug). If we glue $g$ many of them, one gets a \emph{g-handlebody} $\mathcal{H}_g$. The surface that bounds $\mathcal{H}_g$ is called an orientable surface of genus $g$, denoted by $\Sigma_g$.For a $3$-ball $g=0$.
	\item Thickened $2$-sphere (obtained by cutting out a smaller $3$-ball inside a $3$-ball). Similarly, one can remove $n$-many disjoint $3$-balls from a larger $3$-ball.
	\item Thickened $\Sigma_g$.
\end{itemize} 

\noindent All other 3d shapes in $\mathbb{R}^3$ can be obtained by combining the above three types.

\begin{figure}[ht]
	\centering
	\includegraphics[width=10cm]{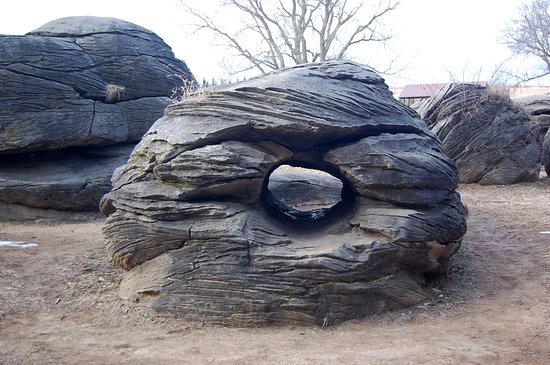}
	\caption{A doughnut shaped rock. Image source : https://sl1nk.com/ktd5797}
	\label{fig8}
\end{figure}

 \noindent Given a convex 3D mosaic, If one does not distinguish between regular and irregular cells, then the topology of the underlying $2$-dimensional complex $X^{(2)}$(consisting of the fracture planes) restricted by the topology of the $3D$ object. For a convex mosaic of a $3$-dimensional solid body $X$, its Euler characteristic is defined by $\chi(X) = c_0 - c_1 + c_2 - c_3$, where $c_i = \#\{i-\textrm{faces in X}\}$. For example, when $X = \mathcal{H}_g$, we have $c_0 - c_1 + c_2 - c_3 = 1-g$.  

\noindent It is also known from basic algebraic topology that $\chi(X^{(2)}) = b_0 - b_1 + b_2$, where $b_i$ ($\geq 0$) is the rank (called $i$th \emph{betti number}) of the $i$th homology group $H_i(X^{(2)}; \mathbb{R})$. For non-trivial convex mosaics $b_2(X^{(2)})$ is always a positive integer, that captures the number of $2$-dimensional holes in $X^{(2)}$. Now, if one only considers the subset $X^{(2)}_{ir}$of $X^{(2)}$, consisting of irregular $2$-faces, then the betti numbers could be drastically different. In particular, it is possible that some of the betti numbers vanish. The following result of Forman is very useful in detecting that.

\vspace{0.15cm}

\noindent \textbf{Theorem (Forman \cite{For}) :} If in a regular CW complex $V$, $\mathcal{F}_p(\alpha^p) > 0$ for all $p$-cells, then $H_p(V;\mathbb{R}) = 0$.

\vspace{0.15cm}

\noindent The computation of Forman-Ricci curvature is computationally more efficient than computing homology group ($N$ vs $N^3$). Comparing the regular/irregular Forman--Ricci curvatures of different $3$D mosaics should give a way to measure the difference in topological regularity.

\section*{Discussion and further directions}

We introduced a version of the Forman--Ricci curvature in the context of irregular convex mosaics. The key point was to define the regular degree of a generic node in an irregular mosaic as the number of straight looking edges (i.e. edges associated to transversely intersecting straight lines). The regular (irregular) curvatures are then defined by replacing usual nodal degrees by regular (irregular) nodal degrees in the formula of Forman--Ricci curvature. We compared the node-based regular curvature distribution and the edge-based Forman--Ricci curvature distribution to distinguish irregular mosaics appearing in nature.

\noindent Our definition of irregular Forman--Ricci curvature can be generalized to non-convex mosaics. Moreover, it can be defined for networks or mosaics on more general types of surfaces and $3$-dimensional solids. This opens up applicability to various mosaic structures appearing in nature. Also, a careful context-based implementation of the weighted-version of regular/irregular Forman--Ricci curvature (both in 2D and 3D) has potential for many diverse applications, and we wish to take up such studies in a future endeavour.

\section*{Author contributions statement}

All authors contributed equally.  All authors reviewed the manuscript. 

\section*{Competing interests} The authors declared that they have no conflicts of interest to
this work.

\end{document}